


\input jytex
\typesize=12pt         \baselinestretch=1200

\def\symbols#1{\ifcase#1 $\bullet$\err@badcountervalue
	\or*\or\markup\dag\or\markup\ddag\or\markup\S%
	\or**\or\markup{\dag\dag}\or\markup{\ddag\ddag}\or\markup{\S\S}%
	\else	$\bullet$\err@badcountervalue
	\fi}
\footnotenumstyle{symbols}

\sectionstyle{center}  \sectionnumstyle{arabic}
\subsectionstyle{left}  \subsectionnumstyle{blank}
\def\sec#1{{\bigfonts\bf\section{#1}}}  \def\subsec#1{{\bf\subsection{#1}}}
\def\eq#1{\eqno\eqnlabel{#1}$$}  \def\ref#1{\markup{[\putref{#1}]}}
\def\puteq#1{eq.~(\puteqn{#1})}  

               \def\tphi{\tilde\phi}
\def\bra#1{\langle{#1}|}          \def\ket#1{|{#1}\rangle}
\def\db#1#2{{d^{#1}{#2}\over (2\pi)^{#1}}}
\def\del{\partial}
\def\dfo{\Delta F_1}		  \def\df1r{\Delta F_{1,{\rm ren}}}
\def\dft{\Delta F_T}
\def\Seff{S_{\rm eff}}
\def\fct{F^{\rm ct}}
\def\L1{{\cal L}^{(1)}}
\def\Leffr{\Lambda_{\rm eff}(r)}
\def\Leff0r{\Lambda_{\rm eff}^0(r)}
\def\Lagr{{\cal L}}                        \def\Lct{{\cal L}_{\rm ct}}
\def\Lwkb{{\Lambda_{\rm WKB}}}
\def\prop0{\Delta_0}	  \def\Om{\Omega}
\def\Tr{{\rm Tr}}
\def\w{\omega}


\pagenumstyle{blank}
\hbox to\hsize{\footnotesize\baselineskip=12pt
  \hfil\vtop{\hbox{\strut CALT-68-1903}
   \hbox{\strut hep-ph/9311369} \hbox{\strut}
   \hbox{\strut DOE RESEARCH AND} \hbox{\strut DEVELOPMENT REPORT}}}
\vskip1.in
\centertext{\baselinestretch=1000\bigsize\bf
Evaluation of the One-loop Effective Action at\\Zero and Finite
Temperature%
\footnote{Work supported in part by the U.S. Dept. of Energy under Contract
  No. DE-FG03-92-ER40701.}}
\vskip .25in 
\centertext{Clarence L. Y. Lee}%
\vskip .2in 
\centertext{\it California Institute of Technology, Pasadena, CA 91125}

\vskip .7in
\centerline{\bf Abstract}
\medskip {\narrower \baselinestretch=1100
A method for determining the leading quantum contributions to the
effective action for both zero and finite temperatures is presented. While
it is described in the context of a scalar field theory, it can be
straight-forwardly extended to include fermions. An extrapolation
procedure which can significantly enhance the computational efficiency is
introduced. This formalism is used to investigate quantum corrections to
the nucleation rate in first-order phase transitions.
 \par}

\vfill\lefttext{October, 1993}\medskip
\newpage \pagenum=0 \pagenumstyle{arabic}

\sec{Introduction}

The evaluation of quantum corrections to classical solutions is an
important problem which pervades much of modern theoretical physics.
However, while effective potentials have been studied extensively,
methods for determining the effective action are less well-developed.
Moreover, the actual evaluation of such effects for
realistic systems has often been hampered by their general intractability
to analytical solution and the lack of efficient computational methods.
\ref{DHN,Was,LiP,Baa,BaK}

In the effective potential approximation to the effective action, quantum
fluctuations are integrated out about a constant classical field --- but
this is not expected to be adequate because the classical field is
generally an inhomogeneous
configuration. The derivative expansion\ref{Cha} improves on this by
accounting for spatially varying background fields; its leading term is
the effective potential. The expansion is a perturbative approximation
which extracts the dominant contribution of short-distance quantum
effects on long-distance physics. When it converges, it provides an
efficient means for performing calculations. However, when it diverges,
one must often resort to brute-force techniques which entail an
explicit, computationally-intensive evaluation. Furthermore, the
derivative expansion fails whenever the potential $V$ is non-convex
($V'' < 0$) in some region of space, which includes an important class
of perturbatively calculated potentials.\ref{BrL93a} It is clear that a
general method, which is also applicable to such cases, is needed.

In this paper, a method for calculating the quantum effects arising
from the effective action is presented. The next section contains the
general formalism for evaluating the effective action. Section 3
discusses the exact formulation of the computational method as well as
extrapolation techniques which improve its convergence properties. This
formalism is applied to the analysis of phase transitions in Section
4.\ref{BrL93a,BrL93b}

\sec{General Formalism}

Consider a scalar field theory with the Lagrangian density
  $${\Lagr}(\phi) = {1\over 2}\del_\mu \phi \del^\mu \phi - V(\phi),
    \eq{Lagr}
where $V$ is the tree-level potential which has a (classical) vacuum at
$\phi_v$. Since this paper considers 3+1 dimensions exclusively,
renormalizability constrains $V$ to
have no polynomials in $\phi$ of higher power than a quartic.
The classical field $\tphi$ is determined by the equation
  $$\del_\mu \del^\mu \tphi = -V'(\tphi). \eq{eom}
The contribution of one-loop quantum effects to the effective action
can be written as
  $$\Tr\L1 = {i\over 2} \Tr
     \ln \left| {\del^2 +V''(\tphi) \over \del^2 +\mu^2} \right|,$$
where $\mu^2 = V''(\phi_v)$ and the trace runs over space-time
coordinates.\footnote{The trace excludes possible negative and zero modes
of the operator $\del^2 +V''(\tphi)$. When such modes arise, they must be
explicitly removed and treated differently.\ref{BrL93a}}
Part of this trace can be evaluated as
  $\Tr\L1 = \Tr' \int \bra{t}\L1\ket{t} dt$,
where $\Tr'$ runs over the remaining spatial degrees of freedom.
Since this paper deals only with time-independent $\tphi$ fields,
specializing to this case means that states in the energy
basis $\ket{\w}$ are eigenstates of the operator in $\L1$. So inserting a
complete set of such states and performing a partial integration yields
  $$\Tr\L1 = -i\int \Tr'\int \left[{1\over -\w^2 -\nabla^2 +V''(\tphi)}
     -{1\over -\w^2 -\nabla^2 +\mu^2}\right] \w^2 {d\w \over 2\pi} dt.$$
Observe the non-locality of this expression; this generic feature of loop
corrections makes exact analytical treatments difficult.
The remaining trace can be conveniently performed over the eigenstates
of the operators in $\L1$: if $\psi^0_j$ and $\psi_j$ are chosen such that
  $$[-\nabla^2 +\mu^2] \psi^0_j = (\w^0_j)^2 \psi^0_j, \eq{V0ppde}
  $$[-\nabla^2 +V''(\tphi)] \psi_j = (\w_j)^2 \psi_j, \eq{Vppde}
where the subscript $j$ indexes the eigenstates, then
  $$\Tr\L1 = -{1\over 2} \int \sum_j (\w_j - \w^0_j) dt.$$
Hence the one-loop effective action can be written as
  $$\Seff(\tphi) = \int \left[\Lagr(\tphi) -\Lagr(\phi_v)\right] d^4x
     +\int \left[ -{1\over 2}\sum_{\w^0_j < \Lambda} (\w_j -\w^0_j)
     +\int\Lct(\tphi,\Lambda) d^3x \right] dt. \eq{Seff}
The bare sum in $\Tr\L1$ is divergent; it is regulated in \puteq{Seff} by
a momentum cut-off $\Lambda$, and a counterterm $\Lct(\tphi,\Lambda)$ has
been added to render it finite.

For time-independent fields $\tphi$, it is more convenient to focus on the
energy $E$ of the system which is related to $\Seff$ through
  $$\Seff(\tphi) = - E(\tphi) \int dt. $$
Then
  $$E(\tphi) = E_{\rm cl}(\tphi) +E^{(1)}(\tphi,\Lambda)
     +E_{\rm ct}(\tphi,\Lambda),  \eq{E}
where $E_{\rm cl}$ is the energy of the classical field configuration
  $$E_{\rm cl}(\tphi) = -\int \left[\Lagr(\tphi) -\Lagr(\phi_v)\right] d^3x,
    \eq{Etree}
$E^{(1)}$ is the one-loop contribution
  $$E^{(1)}(\tphi,\Lambda) =
     {1\over 2}\sum_{\w^0_j < \Lambda} (\w_j -\w^0_j), \eq{E1}
and $E_{\rm ct}$ is the energy due to the one-loop counterterms
  $$E_{\rm ct}(\tphi,\Lambda) = -\int \Lct(\tphi,\Lambda) d^3x.  \eq{Ect}
At finite temperature $T$, the free energy $F$ replaces $E$:\ref{BrL93a}
  $$F_{\rm cl} = E_{\rm cl}, \qquad \dfo=E^{(1)}, \qquad \fct=E_{\rm ct};
    \eq{FEreln}
for bosons, $\tphi(\vec x,\tau)$ is periodic in Euclidean time $\tau$
with period $T^{-1}$, and there is an
additional contribution due to one-loop effects:\ref{BrL93a}
  $$\dft = T \sum_j \ln\left( 1 - e^{-\w_j/T} \over
    1 - e^{-\w_j^0/T} \right) \eq{dftdef}
Observe that no additional counterterms need to be added to $\fct$ because
finite temperatures do not change the short-distance behaviour of the theory.

In the following section, we describe the method developed to evaluate the
the quantum corrections, $\dfo$ and $\dft$, formally given by
\puteq{E1} and (\puteqn{dftdef}).
While for some special situations, the $\w_j$ can be obtained
analytically, this is unfortunately not possible for a general potential
$V''(\tphi)$. Instead the eigenvalues must be found numerically, then for
$\dfo$, the bare sum $\sum_{\w^0_j < \Lambda} (\w_j -\w^0_j)$
is computed explicitly, and finally the counterterm subtracted;
for $\dft$, the sum in \puteq{dftdef} must be performed term-by-term.
To attain reasonable accuracy
this subtraction has to be done at a large cut-off $\Lambda$ (to achieve
convergence) when both the bare sum and the counterterm (which
individually diverge as a function of the cut-off) are numerically very
large. Since the final result is much smaller, each term has to be
determined very precisely, resulting in a heavy computational burden.
Furthermore, the straight-forward approach of evaluating the free energy
by a ``brute-force'' term-by-term summation of the expressions in
\puteq{E1} and (\puteqn{dftdef}) until convergence is reached is
also computationally inefficient.

\sec{Method of Computation}

To circumvent the above-described problem of
having to compute both the regulated bare sum and its counterterm to
very high numerical accuracy, the three-dimensional problem is first
decomposed into channels of definite angular momentum. Then for each
channel, the divergent part of the bare sum is analytically removed
through subtraction with the corresponding divergence in the counterterm,
leaving a much smaller finite piece. Since the contribution of
higher partial-wave channels decrease rapidly, this procedure overcomes
the problem.

An improved computational method is then presented. It is based in part
on the observation that the
higher-energy modes in the spectrum of \puteq{Vppde} are less perturbed
by the potential $V''(\tphi)$ due to the non-uniform background field
$\tphi$ than the lower-energy ones. This allows us to
formulate an approximation method which accounts for the contribution
of the high-energy modes accurately (where the accuracy of the
approximation increases with the energy) so that only some of the
lower-energy modes need to be treated exactly.\ref{Was}

\subsec{Exact Formulation}

The difference in the eigenenergies $\w_j$ and $\w^0_j$ of the unbound
states ($\w > \mu$) can be characterized by the phase shift between the
(asymptotic forms) of the corresponding continuum state eigenfunctions
$\psi_j$ and $\psi^0_j$, as was first shown in one dimension.\ref{DHN}
Since the phase shift is generally a well-behaved, smoothly-varying function
of the energy, it is relatively easy to calculate. Hence it is convenient
to express the free energy in terms of this quantity. To determine the
phase shift, we consider \puteq{Vppde} which determines the fluctuations
about the classical field configuration.

Since most classical solutions $\tphi$ exhibit spherical symmetry
($\tphi = \tphi(r))$, we will
restrict our analysis to such systems. Then the solution to \puteq{Vppde}
can be separated into radial and angular parts by choosing
an eigenfunction of the form
  $$\psi_{nlm}(r,\theta,\phi) = {1 \over r} u_{nl}(r) Y_{lm}(\theta,\phi),
    \eq{eigfn}
where the radial wavefunction is determined by
  $$\left[-{d^2 \over d r^2} + {l(l+1)\over r^2} + V''(\tphi(r)) -
    \w_{n,l}^2\right] u_{n,l}(r) = 0 \eq{ueq}
with the boundary condition $u_{n,l}(0) = 0$. The $Y_{lm}$ are the spherical
harmonics corresponding to a state with total angular momentum $l$ and
$z$-component $m$.

The corresponding equation for $u^0_{n,l}$ where $V''(\tphi)$ is replaced
by $\mu^2$,
  $$\left[-{d^2 \over d r^2} + {l(l+1)\over r^2} + \mu^2 -
    (\w^0_{n,l})^2\right] u^0_{n,l}(r) = 0, \eq{u0eq}
has an exact analytical solution:
  $$u^0_{nl}(r) = \sqrt2 k_n r j_l(k_n r) \eq{usoln}
where $j_l$ is a spherical Bessel function and $k_n^2 = \w_n^2 - \mu^2$.
These solutions have the asymptotic form
  $$u^0_{nl}(r) \rightarrow \sqrt2 \sin(k_n r - {l\pi \over 2}), \>
    r \rightarrow \infty. \eq{u0asymp}
The potentials we consider behave asymptotically as
$V''(\tphi(r)) \rightarrow \mu^2$ when $r \rightarrow \infty$ (which
corresponds to those with finite action). For such potentials, the
asymptotic behaviour of the solution to \puteq{ueq} will be
  $$u_{nl}(r) \rightarrow \sqrt2 \sin(k_n r - {l\pi \over 2}
    + \delta_l(k_n)), \> r \rightarrow \infty. \eq{uasymp}
These equations serve to define the phase shift $\delta_l$ for each
angular momentum channel $l$. Note also that both $u_{n,l}$ and
$u^0_{n,l}$ are $(2l+1)$-fold degenerate.

To facilitate the counting of states, it is convenient to discretize
the eigenvalue spectrum. This procedure can be achieved by
enclosing the system in a box of radius $L$ (where $L$ is
much greater than the range of the potential $V''$) and imposing
the boundary condition
  $$u_{n,l}(L) = 0 \eq{uBC}
which requires that
  $$k_n L - {l\pi \over 2} + \delta_l(k_n) = n\pi. \eq{phase}
Note that such a discretization is implicit in the formal sums in
\puteq{E1} and (\puteqn{dftdef}). The values attained by $\w^0$
(before discretization) as defined by the energy eigenvalue of
\puteq{u0eq} is a continuous spectrum ranging from an energy of $\mu$ to
infinity. The corresponding spectrum for $\w$ determined by \puteq{ueq}
will generally consist of some discrete bound states with energies
$\w_j^2 < \mu^2$ and a continuous spectrum with energies $\w_j^2 > \mu^2$.
The difference in structure between the continuum spectra of the two
systems manifests in a difference in the respective density of states.
Hence it is appropriate to express the sum over eigenenergies for the
states in the continuum as an integral over the density of states:
  $$\sum_j \w^0_j = \sum_l (2l+1)\int_\mu^\infty \w n_l^0(\w) d\w \eq{w0sum}
  $$\sum_j \w_j = \sum_{\w_{nl}^2 < \mu^2} (2l+1)\w_{nl} +
                   \sum_l (2l+1)\int_\mu^\infty \w n_l(\w) d\w \eq{wsum}
where $(2l+1) n_l(\w)$ is the density of states of angular momentum $l$
for the potential $V''(\tphi)$ with an analogous definition for $n_l^0$.
On taking the continuum limit ($L \rightarrow \infty$), it follows from
\puteq{phase} that the densities of
states are related to the continuum phase shift through
  $$n_l(\w) = n_l^0(\w) + {1 \over \pi}{d\delta_l(\w) \over d\w}.\eq{denst}
Now if \puteq{Vppde} has $N$ bound states, then since \puteq{Vppde} and
\puteq{V0ppde} must have the same total number of states,
  $$N + \sum_l (2l+1)\int_\mu^\infty n_l(\w) d\w =
    \sum_l (2l+1)\int_\mu^\infty n_l^0(\w) d\w. \eq{numst}
For a finite potential, this implies $N \pi = \delta(\mu)$.

It is convenient to define the free energy in each angular momentum
channel such that
  $$\dfo(\Lambda) = \sum_l (2l+1) \dfo^l(\Lambda), \eq{df1ldef}
  $$\dft = \sum_l (2l+1) \dft^l, \eq{dftldef}
and to similarly partition the counterterm energy as
  $$\fct(\Lambda) = \sum_l (2l+1) \fct_l(\Lambda), \eq{fctldef}
then from the above equations
  $$\dfo^l(\Lambda) = {1\over 2} \sum_{\w_{nl}^2 < \mu^2} (\w_{nl} - \mu)
                      -{1\over 2\pi}\sum_l \int_\mu^\Lambda
                       \delta_l(\w) d\w, \eq{df1l}
and
  $$\dft^l = T \sum_{\w_{nl}^2 < \mu^2}
                \ln\left({1-e^{-\w_{nl}/T}\over 1-e^{-\mu/T}}\right)
    -{1\over\pi}\int_\mu^\infty {\delta_l(\w)\over e^{\w/T}-1} d\w. \eq{dftl}

The Appendix discusses the renormalization of the Lagrangian given by
\puteq{Lagr}. It is shown there that the contribution of the
counterterms to the energy are of the general form
  $$\int \left[ g(\w^2 -p^2 -\mu^2) \int h(x) d^3x \right] \db{3}{p}
     {d\w \over 2\pi} \equiv
    \int_{-\infty}^\infty \Tr'(gh) {d\w \over 2\pi}, \eq{ectgen}
where $g$ is a power of the propagator, $h$ is a function of
$\phi$ and its derivatives, and $\Tr'$ is a trace over the spatial
variables. The partial wave decomposition of these contributions is
achieved by taking the trace with respect to the eigenstates of
\puteq{V0ppde} denoted here by $\ket{nlm}$:
  $$\eqalign{\Tr'(gh) =& \sum_{nlm} \sum_{n' l' m'}
                         \bra{nlm} g(\w^2 +\nabla^2 -\mu^2) \ket{n' l' m'}
                         \bra{n' l' m'} h(r) \ket{nlm} \cr
     =& {1\over\pi} \sum_l (2l+1) \int_0^\infty \left[g(\w^2 -p^2 -\mu^2)
                    \int_0^\infty h(r)|u_{pl}^0(r)|^2 dr\right] dp. \cr}
     \eq{trpdef}
{}From the Appendix, the counterterm contribution to the free energy is
  $$\fct(\Lambda) =-{i\over 2}\int_{-\infty}^\infty
            \Tr'\left\{\prop0(\w,p) \left[m^2(r)-\mu^2 \right]
       +{1\over 2}\prop0(\w,p)^2 \left[m^2(r)-\mu^2\right]^2 \right\}
       {d\w\over 2\pi}, \eq{fcttr}
where
  $$\prop0(\w,p) = {1 \over \w^2 -p^2 -\mu^2 +i\epsilon},$$
and $m^2(r) = V''(\tphi(r))$.
Evaluating the trace using \puteq{trpdef} yields
  $$\eqalign{\fct_l(\Lambda) = \int_0^{\Lambda_p} \biggl\{
     -&{1\over 4\pi}{1\over(p^2+\mu^2)^{1/2}}
      \int_0^\infty |u_{pl}^0(r)|^2 \left[m^2(r)-\mu^2 \right] dr \cr
     +&{1\over 16\pi}{1\over(p^2+\mu^2)^{3/2}}
      \int_0^\infty |u_{pl}^0(r)|^2 \left[m^2(r)-\mu^2 \right]^2 dr\biggr\}
    dp,\cr} \eq{fctltr}
where $\Lambda_p = \sqrt{\Lambda^2 - \mu^2}$ is a three-momentum cut-off.

This completes the formulation of the method for the exact calculation of
the free energy. However, as we have
remarked above, the convergence of such an exact computation can be
sufficiently slow so that extrapolation techniques can be useful.
Amongst the various such procedures, we consider in particular the WKB
approximation, which provides an analytic expression for the phase
shift that is valid at high energies and hence can significantly
reduce the effort required to evaluate the phase shift integral.\ref{Was}

\subsec{WKB-Improved Method}

A differential equation of the form
  $$\left[ {d^2 \over dx^2} + k^2(x) \right] f(x) = 0 \eq{wkbde}
has an approximate WKB solution given by
  $$f_{WKB}(x) = {\exp\left[i\int_0^x k(\w,y) dy\right] \over \sqrt{k(\w,x)}}
    \eq{wkbsoln}
which is valid when the wavelength is much less than the distance scale
over which $k$ varies:
  $${1\over k^2}{dk\over dx} \ll 1$$
where $k(\w,x)$ is the local wavenumber
  $$k(\w,x) = \sqrt{\w^2 - V''(\tphi(x))}.$$
Hence the accuracy of the WKB approximation increases with energy.
The phase shift for such solutions is given by
  $$\delta^{\rm WKB}(\w) = \int_{-\infty}^{\infty}
            \left[ k(\w,x) -\lim_{y\to\infty}k(\w,y) \right]dx. \eq{phwkbdef}
Explicitly,
  $$\delta_l^{\rm WKB}(\w) =
    \int_{a(\w)}^\infty \sqrt{\w^2 -m^2(r) -{l(l+1)\over r^2}} dr -
    \int_{a_0(\w)}^\infty \sqrt{\w^2 -\mu^2 -{l(l+1)\over r^2}} dr, \eq{phwkb}
where $a$ and $a_0$ denote the classical turning points defined by
  $$\w^2 -m^2(a) -{l(l+1)\over a^2} = 0, \qquad {\rm and} \qquad
    \w^2 -\mu^2 -{l(l+1)\over a_0^2} = 0.$$
Applying this method to \puteq{ueq} yields an analytic expression for
the energy integral of the phase shift:
  $$\eqalign{\int_\mu^\Lambda \delta_l^{WKB}(\w) d\w = \int_0^\infty
      \biggl[&\int_{\Om(r)}^\Lambda \sqrt{\w^2-m^2(r)-{l(l+1)\over r^2}}
             \theta(\Lambda-\Om(r)) d\w \cr
            &-\int_{\Om_0(r)}^\Lambda \sqrt{\w^2-\mu^2-{l(l+1)\over r^2}}
             \theta(\Lambda-\Om_0(r)) d\w \biggr] dr \cr} \eq{wkbphint}
with
  $$\Om(r) = \sqrt{m^2(r) + {l(l+1)\over r^2}}, \qquad
    \Om_0(r) = \sqrt{\mu^2 + {l(l+1)\over r^2}}, $$
and $\theta(x)$ is the unit step-function.
Observe that since the high-energy behaviour of the phase shift is
independent of the angular momentum, the energy of each angular
momentum channel is logarithmically divergent:
  $$\dfo^l(\Lambda) = {1\over 4\pi} \ln\left(\Lambda\over\mu\right)
             \int_0^\infty \left[m^2(r)-\mu^2 \right] dr
            + {\cal O}(\Lambda^0) \eq{df1he}
Now the divergent piece in $\dfo^l$ can be analytically combined with the
infinite part of $\fct_l$ in \puteq{fctltr} to leave only finite terms.
Performing this subtraction and taking the limit
$\Lambda \rightarrow \infty$ gives the final expression for the
WKB-improved, temperature-independent renormalized free energy:
  $$\eqalign{
    \df1r^l =& \lim_{\Lambda\to\infty}[\dfo^l(\Lambda) +\fct_l(\Lambda)]\cr
     =&{1\over 2}\sum_{\w_{nl}^2<\mu^2} (\w_{nl}-\mu)
      -{1\over 2\pi} \int_\mu^\Lwkb \delta_l(\w) d\w \cr
     &-{1\over 2\pi} \int_0^\infty \left\{\chi_l(\Lwkb,r)
      +\kappa_l(r) \left[m^2(r)-\mu^2 \right]
      +\rho_l(r) \left[m^2(r)-\mu^2 \right]^2 \right\} dr \cr}
   \eq{df1lwkb}
In this equation, $\chi_l$ is the contribution from the WKB phase shift
above $\Lwkb$,
  $$\eqalign{
    \chi_l(r) =& {m_l^0(r)^2 -m_l(r)^2 \over 4}
                -{1\over 2}\Leffr \sqrt{\Leffr^2 -m_l(r)^2} \cr
               &+{1\over 2}\Leff0r \sqrt{\Leff0r^2 -m_l^0(r)^2} \cr
     &+{1\over 2}m_l(r)^2 \ln{\Leffr +\sqrt{\Leffr^2-m_l(r)^2}\over\mu}\cr
     &-{1\over 2}m_l^0(r)^2 \ln{\Leff0r +\sqrt{\Leff0r^2-m_l^0(r)^2}
                 \over\mu}, \cr} \eq{chil}
where
  $$m_l^0(r)^2 = \mu^2 + {l(l+1)\over r^2}, \qquad
    m_l(r)^2 = m^2(r) + {l(l+1)\over r^2},$$
  $$\Leff0r = {\rm max} \left(\Lwkb,\Om_0(r)\right), \qquad
    \Leffr = {\rm max} \left(\Lwkb,\Om(r)\right),$$
and $\Lwkb$ denotes the energy above which the phase shift is
computed by the WKB method. The remaining terms in the last integral come
from finite parts of the counterterm with
  $$\kappa_l(r) = \int_0^\infty {s^2|j_l(s)|^2 -{1\over 2}\over
                               \sqrt{s^2 +(\mu r)^2}} ds, \eq{kappa}
and
  $$\rho_l(r) = -{r^2 \over 4}\int_0^\infty
     {s^2 |j_l(s)|^2 \over (s^2 + (\mu r)^2)^{3/2}} ds. \eq{rho}
Equation (\puteqn{df1lwkb}) indicates that $\df1r^l$ can now be computed
by first summing over the bound state energies, then the continuum
state contributions can be evaluated by explicitly computing the exact
phase shift only up to $\Lwkb$, beyond which the WKB method provides
an analytical expression that accounts for contributions at higher
energies. Note that while the WKB procedure entails an approximation,
its accuracy can be made such that the difference between the exact and
the WKB results is smaller than the desired precision. Finally, summation
over $l$ yields
  $$\df1r = \sum_l (2l+1) \df1r^l. \eq{df1r}

Since $\dft$ is not divergent, it can be computed exactly using
\puteq{dftldef} and (\puteqn{df1l}), or by replacing the exact phase shift
$\delta_l$ above a certain energy scale by the approximate WKB phase shift
$\delta_l^{\rm WKB}$ given by \puteq{phwkb}.

\sec{Application and Discussion}

In ref.[\putref{BrL93a}] these methods have been used to calculate the
free energy of an instanton configuration which determines the decay rate
in a first order phase transition. The
computation of $\dfo$ will be described first. It is found that the
accuracy available on conventional computers prevents a precise
determination of this quantity when it is straight-forwardly evaluated
as in \puteq{E1} --- that is, by doing the bare sum and subtracting
the counterterm, without a decomposition into partial waves. When $\dfo$
is computed exactly, by utilizing such a decomposition, very high numerical
accuracy is still required because for each $l$ the bare sum and
$\fct_l(\Lambda)$ must be evaluated at a large value of the cut-off
$\Lambda$. But since both quantities diverge as a function of $\Lambda$,
we find that convergence with reasonable accuracy is still difficult
to attain. In contrast, evaluation of $\df1r$ using the WKB-improved
method consisting of \puteq{df1lwkb} and (\puteqn{df1r}) converges
rapidly for much lower values of the cut-off $\Lwkb$ and typically only
the first fifty partial waves need to be summed; the parameters required
for convergence are very much dependent on the nature $V''(\tphi(r))$ and
the values we have quoted come from the potentials we have examined.

The exact computation of $\dft$ can be performed by evaluating
\puteq{dftldef} and (\puteqn{dftl}), but at high temperatures it is found
that several
hundred partial waves must be summed to attain convergence. When the
exact phase shift is replaced by the approximate WKB expression at high
energies, there is a reduction in the computational burden and
the same number of angular momentum channels must be summed. The
improvement is not marked as it was for $\df1r$ in part because $\dft$ is
not renormalized. The results of these computations are summarized in
Tables 4 and 5 of ref.[\putref{BrL93a}].

In summary, we have elucidated a method for the exact evaluation the
effective action to one-loop. The WKB extrapolation scheme was devised
to reduce the computational effort. These methods enable an
efficient calculation of the free energy associated with a phase
transition, as detailed above. However, the applicability of this
method is not limited to this example. Rather, it can be utilized in a
broader variety of problems involving the non-perturbative evaluation of
observables in a non-uniform background in quantum field theory\ref{Raj}
as well as in classical systems\ref{DoL}. It can also be generalized to
encompass theories with fermions.\ref{Lee}

\sectionnumstyle{blank}
\newpage
\sec{Appendix: One-loop Renormalization of the Scalar Field Theory}

This Appendix discusses the one-loop and renormalization
of the scalar field theory described by \puteq{Lagr}.
The classical vacuum $\phi_v$ satisfies
  $$V'(\phi_v) = 0 \ {\rm and}\ \mu^2 = V''(\phi_v) > 0. \eq{phifdef}
At one-loop the only divergent graphs are those with one and two vertices
corresponding to quadratic and logarithmic divergences, respectively.

It is convenient to adopt a renormalization scheme where the counterterms
are chosen to exactly cancel the divergent graphs as shown in
Figs. 1 and 2. These conditions are imposed at
zero external momenta; this choice has the advantage that the one-loop
contribution to the effective potential $V_1$ satisfies
  $$V_1(\phi_v) = V'_1(\phi_v) = V''_1(\phi_v) = 0, \eq{rencond}
so that \puteq{phifdef} is unchanged at one-loop. Then the counterterm
Lagrangian to be added to \puteq{Lagr} is
  $${\cal L}_{\rm ct} = \alpha \left[V''(\phi) -\mu^2 \right]
     +{1\over 2}\beta \left[V''(\phi) -\mu^2 \right]^2,   \eq{Lct}
where
  $$\alpha = {1\over 2}\int {i\over k^2-\mu^2+i\epsilon} \db{4}{k},$$
and
  $$\beta = {1\over 2}\int {i\over (k^2-\mu^2+i\epsilon)^2} \db{4}{k}.$$
The terms in \puteq{Lct} involving $\alpha$ and $\beta$ renormalize the
graphs with one and two external vertices, respectively. These
divergent integrals can be suitably regularized by imposing a
momentum cut-off $\Lambda$.

\sec{Acknowledgements}

The author wishes to thank David Wasson and David Brahm for
enlightening discussions.


\newpage
\sec{References:}

\def\pr#1{{\it Phys.\ Rev.} {\bf #1}}
\def\prl#1{{\it Phys.\ Rev.\ Lett.} {\bf #1}}
\def\zp#1{{\it Z.\ Phys.} {\bf #1}}
\baselinestretch=1000
\begin{putreferences}
\reference{Raj}{A detailed discussion can be found in R. Rajaraman,
   {\it Solitons and Instantons}, North-Holland Press, Amsterdam, 1982).}
\reference{DoL}{For a survey, see {\it Phase Transitions and Critical
  Phenomena}, (Academic Press, London, 1983), Vol. 8, edited
  by C. Dom \& J. L. Lebowitz.}
\reference{BrL93a}{D. E. Brahm \& C. L. Y. Lee, CALT-68-1881 (Nov.\ 1993).}
\reference{BrL93b}{D. E. Brahm \& C. L. Y. Lee, work in progress.}
\reference{Cha}{See for example, L.-H. Chan, \prl{54}:1222 (1985);
                \prl{56}:404(E) (1986).}
\reference{Was}{D. A. Wasson, Ph.D. thesis, California Institute of
  Technology, 1990.}
\reference{Baa}{J. Baacke, \zp{C47}:263 (1990).}
\reference{BaK}{J. Baacke \& V. G. Kiselev, DO-TH-93/18.}
\reference{DHN}{R. F. Dashen, B. Hasslacher \& A. Neveu, \pr{D10}:4130
  (1974).}
\reference{Lee}{C. L. Y. Lee, work in progress.}
\reference{LiP}{M. Li \& R. J. Perry, \pr{D37}:1670 (1988).}
\end{putreferences}

\newpage
\subsec{Figure Captions}

\item{Figure 1.}
  Renormalization scheme for the divergent one-loop graphs with one
  vertex. A box with a cross denotes a counterterm insertion.
\medskip
\item{Figure 2.}
  Renormalization scheme for the divergent graphs at one-loop order with
  two vertices.

\bye